\documentclass[sigconf, nonacm]{acmart}
\usepackage{graphicx}
\usepackage{balance}
\usepackage{algorithm}
\usepackage[noend]{algpseudocode}
\usepackage{subcaption}

\usepackage{listings}
\definecolor{mygreen}{rgb}{0,0.6,0}
\definecolor{mygray}{rgb}{0.5,0.5,0.5}
\definecolor{mymauve}{rgb}{0.58,0,0.82}
\definecolor{TUMBlau}{RGB}{0,101,189} 
\definecolor{TUMBlauDunkel}{RGB}{0,82,147} 
\definecolor{TUMBlauHell}{RGB}{152,198,234} 
\definecolor{TUMBlauMittel}{RGB}{100,160,200} 
\definecolor{TUMElfenbein}{RGB}{218,215,203} 
\definecolor{TUMGruen}{RGB}{162,173,0} 
\definecolor{TUMGruenHell}{RGB}{188,207,30} 
\definecolor{TUMGruenDunkel}{RGB}{22,164,00} 
\definecolor{TUMGruenDunkel2}{HTML}{169100} 
\definecolor{TUMRosa}{RGB}{227,130,143} 
\definecolor{TUMRosaHell}{RGB}{242,144,149} 
\definecolor{TUMOrange}{RGB}{243,145,0} 
\definecolor{TUMOrangeHell}{RGB}{247,166,0} 
\definecolor{TUMSenf}{RGB}{202,171,41} 
\definecolor{TUMSenfHell}{RGB}{232,200,55} 
\definecolor{TUMGrau}{gray}{0.6} 
\usepackage{verbatim}
\usepackage{amsmath}
\newcommand{\nat}{\mathbb{N}}

\newcommand{\Epsilon}{\mathcal{E}}

\newcommand{\abs}[1]{\ensuremath{\lvert#1\rvert}}
\newcommand{\mbeq}{\overset{!}{=}}
\newcommand{\half}{\frac{1}{2}}
\usepackage{fancyvrb}
\DefineVerbatimEnvironment{Code}{BVerbatim}{baseline=t}
\usepackage{tikz}
\usetikzlibrary{arrows,positioning, automata, calc, fit, shapes.geometric, decorations.pathreplacing, decorations.pathmorphing, spy, shapes}
\usepackage{pgfplots}
\usepackage{pgfplotstable}
\usepgfplotslibrary{groupplots}
\usepackage{tikz-qtree}
\usepackage{multicol}

\newcommand\vldbavailabilityurl{https://github.com/tum-db/fastfactorizedlearning}

\begin{document}
\title{Fast Factorized Learning}
\subtitle{Powered by In-Memory Database Systems}


%
%


\author{Bernhard Stöckl}
\affiliation{\institution{Technical University of Munich}}
\email{bernhard.stoeckl@tum.de}
\author{Maximilian E. Schüle}
\email{maximilian.schuele@uni-bamberg.de}
\affiliation{\institution{University of Bamberg}}

\begin{abstract}
Learning models over factorized joins avoids redundant computations by identifying and pre-computing shared cofactors.
Previous work has investigated the performance gain when computing cofactors on traditional disk-based database systems.
Due to the absence of published code, the experiments could not be reproduced on in-memory database systems.

This work describes the implementation when using cofactors for in-database factorized learning.
We benchmark our open-source implementation for learning linear regression on factorized joins with PostgreSQL---as a disk-based database system---and HyPer---as an in-memory engine.
The evaluation shows a performance gain of factorized learning on in-memory database systems by 70\% to non-factorized learning and by a factor of 100 compared to disk-based database systems.
Thus, modern database engines can contribute to the machine learning pipeline by pre-computing aggregates prior to data extraction to accelerate training.
\end{abstract}

\maketitle

%

\section{Introduction} \label{sec:Intro}

Machine learning models are typically trained on a single, unnormalized data set.
Without normalization, functional dependencies between attributes lead to data replication within the final table.
This results in redundant computations when training a model.

Identifying a model's common cofactors reduces redundancy and training time.
Cofactors are shared factors between aggregations like sums over grouped data.
Cofactors can be precomputed on normalized data if the original query plan is known.
Schleich et~al.~\cite{factor, layer,learningLinear} use functional dependencies for in-database training of linear regression.
They speed up gradient descent on PostgreSQL as a disk-based database system by pre-materializing cofactors.

This study elaborates on the performance gain of in-memory database systems for factorized learning~\cite{DBLP:journals/pvldb/HuangSL023}.
Due to the lack of original source code, we provide our own open source implementation for computing cofactors for gradient descent on top of database systems\footnote{\url{https://github.com/tum-db/fastfactorizedlearning}}.
We will use this implementation to compare the performance of PostgreSQL and HyPer for in-database factorized learning.

\begin{figure}[tb]
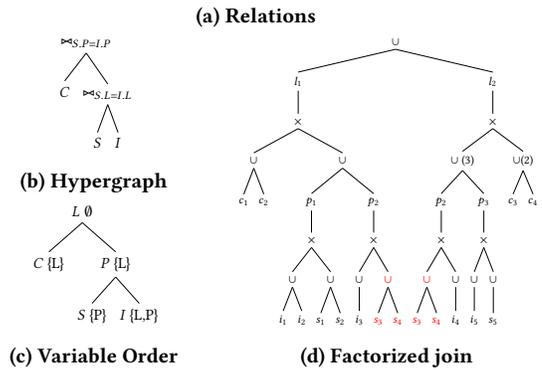

\centering
\begin{subfigure}[t]{\linewidth}
\centering
\scriptsize
\begin{tabular}{cc}
\multicolumn{2}{c}{\textbf{S}ales}\\
P & S \\\hline
$p_1$ & $s_1$\\
$p_1$ & $s_2$\\
$p_2$ & $s_3$\\
$p_2$ & $s_4$\\
$p_3$ & $s_5$\\
\end{tabular}
\begin{tabular}{ccc}
\multicolumn{3}{c}{\textbf{I}nventory}\\
L & P & I \\\hline
$l_1$ & $p_1$ & $i_1$ \\
$l_1$ & $p_1$ & $i_2$ \\
$l_1$ & $p_2$ & $i_3$ \\
$l_2$ & $p_2$ & $i_4$ \\
$l_2$ & $p_3$ & $i_5$ \\
\end{tabular}
\begin{tabular}{cc}
\multicolumn{2}{c}{\textbf{C}ompetition}\\
L & C \\\hline
$l_1$ & $c_1$ \\
$l_1$ & $c_2$ \\
$l_2$ & $c_3$ \\
$l_2$ & $c_4$ \\\\
\end{tabular}
\begin{tabular}{ccccc}
\multicolumn{5}{c}{$C \bowtie (S \bowtie I)$}\\
L & C & P & I & S\\\hline
\color{red} $l_1$ &\color{red}  $c_1$ & \color{red} $p_1$ & \color{blue} $i_1$ & \color{orange}  $s_1$\\
\color{red} $l_1$ &\color{red}  $c_1$ & \color{red} $p_1$ & \color{blue} $i_1$ & \color{violet}  $s_2$\\
\color{red} $l_1$ &\color{red}  $c_1$ & \color{red} $p_1$ & \color{pink} $i_2$ & \color{orange}  $s_1$\\
$l_1$ & $c_1$ & $p_2$ & $i_3$ & $s_3$\\
\multicolumn{5}{c}{\dots}\\
\end{tabular}
\caption{Relations}
\label{fig:sale_data}
\end{subfigure}%

\begin{subfigure}[t]{0.4\linewidth}
\centering
\scalebox{.65}{
\Tree
[.{$\bowtie_{S.P=I.P}$}
 {$C$}
        [.{$\bowtie_{S.L=I.L}$} $S$ $I$ ]]
}
\caption{Hypergraph}
\label{fig:sale_hyper}
\centering
\scalebox{.65}{
\Tree
[.{$L$~$\emptyset$}
 {$C$~\{L\}}
        [.{$P$~\{L\}} $S$~\{P\} {$I$~\{L,P\}} ]]
}
\caption{Variable Order}
\label{fig:sale_vo}
\end{subfigure}
\begin{subfigure}[t]{0.5\linewidth}
\centering
\centering
\scalebox{.5}{
\Tree [.{$\cup$}
[.{$l_1$}
[.{$\times$}
[.$\cup$ $c_1$ $c_2$ ]
[.$\cup$
[.$p_1$ [.{$\times$} [.$\cup$ $i_1$ $i_2$ ] [.$\cup$ $s_1$ $s_2$ ] ]]
[.$p_2$ [.{$\times$} [.$\cup$ $i_3$ ] [.\color{red}$\cup$ \color{red}$s_3$  \color{red}$s_4$  ] ]]]]]
[.$l_2$
[.{$\times$}
[.{$\cup$ (3)}
[.$p_2$ [.{$\times$ } [.\color{red}$\cup$ \color{red}$s_3$  \color{red}$s_4$  ] [.$\cup$ $i_4$ ] ]]
[.$p_3$ [.{$\times$ } [.$\cup$ $i_5$ ] [.$\cup$ $s_5$ ] ]]]
[.$\cup$(2) $c_3$ $c_4$ ]]]]
}
\caption{Factorized join}
\label{fig:sale_fact}
\end{subfigure}
\caption[Different representations of the relations $Sales$, $Inventory$ and $Competition$ and their join]{$(a)$ Relations: Sales(Product, Sale), Inventory(Location, Product, Inventory), Competition(Location, Competitor),
$(b)$ Hypergraph of the natural join,
$(c)$ Variable order of the natural join,
$(d)$ Factorized join over the given schema (taken from \cite[Figure~1]{factor})}
\label{fig:sale}
\end{figure}


This paper is organized as follows:
Section \ref{sec:fact} explains how calculating aggregates for factorization reduces redundancy.
Section \ref{sec:linReg} combines linear regression with factorization for in-database gradient descent.
Section \ref{sec:impl} describes the implementation that was created using these concepts along this work.
Section \ref{sec:Eval} then compares the implementation on top of PostgreSQL \cite{psql} with HyPer \cite{hyper}.
The last section concludes this study with an outlook on other models that benefit from factorization.

\section{Factorization} \label{sec:fact}
This section describes the idea behind factorization and how different operations can be computed on a factorized join.

\subsection{Concept and Advantages of Factorized Joins} \label{sec:concept}
Usually, the result of a join contains some redundancy (cf.~\autoref{fig:sale_data}).
Factorization tries to avoid duplicates by representing the join symbolically.
For example the product $p_1$ combines $s_1$ and $s_2$ in $Sales$ with $i_1$ and $i_2$ in $Inventory$, which can be represented as $\{i_1, i_2\}\times \{s_1,s_2\}$.
Doing the same for $p_2$ results in $\{i_3\} \times \{s_3,s_4\}$. Since $l_1$ is associated both with $p_1$ and $p_2$ it can be represented as the union of the two previous results. Doing this for all entries results in the factorized representation shown in \autoref{fig:sale_fact}.

Since $p_2$ is always associated with both $s_3$ and $s_4$, independent of the values of location and competitor, the union of those two elements can be cached.
These valued are shown in red in \autoref{fig:sale_fact} through which $p_2$ references $S_{34}=\{s_3,s_4\}$.
This is possible because $S$ is conditionally independent of $L$ given $P \circ S$ and $L$ do not appear in the same relation \cite[Definition 3.2]{factor}. 

In general, a join query $Q$ over a database $D$ results in a flat representation of size $O(\abs{D}^{\rho^*(Q)})$ and a factorized representation of size $O(\abs{D}^{fhtw^*(Q)})$ \cite[Theorem 3.4]{factor}. The measure $\rho^*(Q)$ represents the fractional edge cover number which is defined as the minimal number of edges needed to cover all nodes in a hypertree \cite{hypertree}. A hypertree is a tree in which every node represents the union of some hyperedges and all subtrees rooted at $p$ intersected with the hyperedges used in $p$ are included in the set of $p$. Applied to databases this means we ask for the minimal number of relations taking part in $Q$ that are required to cover all attributes in $Q$. $fhtw^*(Q)$, on the other hand, stands for the fractional hypertree width and is defined as the width of a tree decomposition $T$ of the hypergraph of $Q$ such that every node of the tree represents a subset $S$ of the relations in $Q$ and these additional properties are fulfilled \cite{hypertree,treewidth}:
\begin{itemize}
\item Every relation $R$ in $Q$ is included in at least one node of the tree: $\forall R\in Q: \exists S \in T : R \in S$
\item All nodes on the simple path connecting two nodes that contain the same attribute $a$ contain this attribute as well:\\ $\forall S,V,E \in T: \forall a \in \bigcup_{R \in Q}:\\ \left( a \in S \land a \in V \land E \in path(S,V) \implies a \in E \right)$
\item $T$ is a tree decomposition with minimal width. The width $w$ of $T$ is defined as the maximum number of relations contained in a single node of $T$:
 $w = max\left\{ \abs{S} \mid S \in T \right\}$
\end{itemize}
With those definitions $1 \leq fhtw^*(Q) \leq \rho^*(Q) \leq \abs{Q} $ holds and the gap between both of them can be as large as $\abs{Q}$ \cite[p.8]{factor}.

\subsection{Variable Orders} \label{sec:vaOr}
Given a join query $Q$, a variable order $\Delta$ for this query is a pair $(T, key)$ such that:
\begin{itemize}
\item ``$T$ is a rooted forest with one node per variable in $Q$ such that the variables of each relation symbol in $Q$ lie along the same root-to-leaf path in T'' \cite[Definition 3.2]{factor}
\item ``$key$ maps each variable $A$ to the subset of its ancestors in $T$ on which the variable in the subtree rooted at A depend'' \cite[Definition 3.2]{factor}
\end{itemize}
\noindent
The function $key$ is used to encode the independence of variables in the variable order.
The key always contains the parent of the current node: their direct connection implies that they are in the same relation.

Nodes on different paths never depend on each other:
First, per definition, all nodes that are in the same relation have to be in the same root-to-leaf path in T.
Second, variables occurring in the same relation depend on each other \cite[Definition 3.2]{factor}.

Therefore, $key(B) \subseteq key(A)$ holds for all variables A, B such that A is the parent of B in the variable order $\Delta$.
\autoref{fig:sale_vo} gives an example of a variable order corresponding to the factorized join depicted in \autoref{fig:sale_fact}:
$L$ is the root of the tree and, therefore, has no variables it depends on.
Its children $C$ and $P$ both depend on $L$ which is denoted by the set $\{L\}$.
$P$ also has two children which depend on it, but only $I$ depends on $L$ as well, since there is no relation containing both $L$ and $S$.

\subsection{Aggregates on Factorized Joins} \label{sec:operations}
Given a join, SQL aggregates can be computed in one pass over its factorized representation by replacing the unions, cross products and values with functions corresponding to the required aggregate.
\begin{figure}[tb]
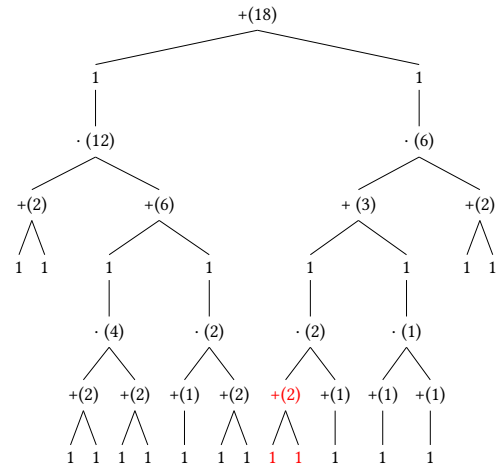

\begin{minipage}{\linewidth}
\centering
\scalebox{.8}{
\Tree [.{+(18)}
[.1
[.{$\cdot$ (12)}
[.+(2) 1 1 ]
[.+(6)
[.1 [.{$\cdot$ (4)} [.+(2) 1 1 ] [.+(2) 1 1 ] ]]
[.1 [.{$\cdot$ (2)} [.+(1) 1 ] [.+(2) 1 1 ] ]]]]]
[.1
[.{$\cdot$ (6)}
[.{+ (3)}
[.1 [.{$\cdot$ (2)} [.\color{red}+(2) \color{red}1 \color{red}1 ] [.+(1) 1 ] ]]
[.1 [.{$\cdot$ (1)} [.+(1) 1 ] [.+(1) 1 ] ]]]
[.+(2) 1 1 ]]]]
}
\caption{Counting all elements from join in \autoref{fig:sale}}
\label{fig:count}
\end{minipage}
\end{figure}
\begin{figure}[tb]
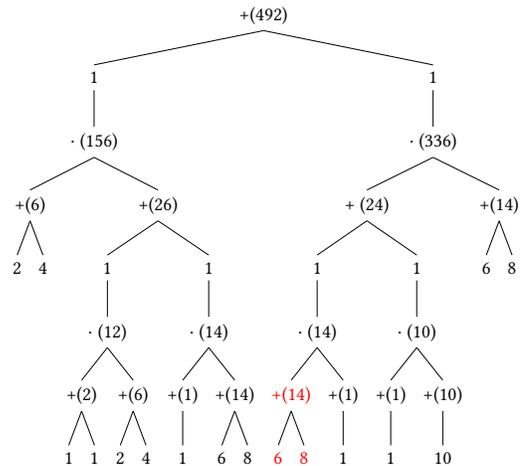

\begin{minipage}{\linewidth}
\centering
\scalebox{.8}{
\Tree [.{+(492)}
[.1
[.{$\cdot$ (156)}
[.+(6) 2 4 ]
[.+(26)
[.1 [.{$\cdot$ (12)} [.+(2) 1 1 ] [.+(6) 2 4 ] ]]
[.1 [.{$\cdot$ (14)} [.+(1) 1 ] [.+(14) 6 8 ] ]]]]]
[.1
[.{$\cdot$ (336)}
[.{+ (24)}
[.1 [.{$\cdot$ (14)} [.\color{red}+(14) \color{red}6 \color{red}8 ] [.+(1) 1 ] ]]
[.1 [.{$\cdot$ (10)} [.+(1) 1 ] [.+(10) 10 ] ]]]
[.+(14) 6 8 ]]]]
}
\caption{Computing $SUM(Sale\cdot Competitor)$}
\label{fig:aggSample2}
\end{minipage}
\end{figure}
For example, if we want to count all rows in the join, we can replace unions by sums, cross products by multiplications and map all values to 1 \cite[p. 2]{factor}.
Applying this to the factorized representation shown in \autoref{fig:sale_fact} results in the tree depicted in \autoref{fig:count}. The values highlighted in red indicate caching.

Another example is the query $SUM(Sale\cdot Competitor)$ which would be computed by replacing cross products with multiplications and unions by sums again. However, this time the values for $Sale$ and $Competitor$ are mapped to their real values while all other values are mapped to 1.
\autoref{fig:aggSample2} shows how this application would affect the factorized representation from \autoref{fig:sale_fact}. Since the sample doesn't provide any actual data for the elements we assume that every element has a value of twice its index (so e.g. $i_1=2$).
The red values indicate caching as in the previous example.

\section{Linear Regression} \label{sec:linReg}
This section explains how linear regression works and how data can be preprocessed with feature scaling to achieve better results.
Section \ref{sec:rewriting} shows how the converging procedure $gradient\ descent$, which is used within linear regression, can be rewritten to support factorized representations.

\subsection{Objective Function} \label{sec:obj}
The goal of linear regression is to find the parameters $\theta_j$ for a linear function $$f(x_1, \dots, x_n) = y = \theta_1 \cdot x_1 + \dots + \theta_n \cdot x_n + c$$ so that it approximates a given training dataset with $m$ rows determined by a join query over a database:
$${(y^{(1)}, x^{(1)}_1, \dots, x^{(1) }_n), \dots, (y^{(m)}, x^{(m)}_1, \dots , x^{(m)}_n)}.$$
This dataset therefore consists of $n$ values $x_j^{(i)}$ which are called features and will be used as input for the function and a label $y^{(i)}$ which is approximated by this function.
For the constant offset $c$, we add another feature which always has the value 1 so that we get the function:
$$h_\theta(x) = \theta_0 + \theta_1 \cdot x_1 + \dots + \theta_n \cdot x_n = \sum_{j=0}^n (\theta_j \cdot x_j)\ \text{ \ \cite[p. 4]{layer}\cite{andrew_1}}$$
An objective function $\Epsilon(\theta)$ evaluates the quality of the values for $\theta$.
A commonly used error function is least squares function: 
$$\Epsilon(\theta) = \half\sum_{i=1}^m\left( h_\theta(x^{(i)}) - y^{(i)} \right)^2 + \lambda R(\theta) \ \text{ \ \cite[p. 3]{learningLinear}\cite[p. 6]{factor}}$$
$R(\theta)$ is called a regularization term weighted with $\lambda$ and is used to avoid overfitting.
Some possible regularization terms are $\lambda\sum_{j=0}^n\theta_j^2$ (Ridge), $\lambda\sum_{j=0}^n\abs{\theta_j}$ (Lasso) and $\lambda_1\sum_{j=0}^n\theta_j^2 + \lambda_2\sum_{j=0}^n\abs{\theta_j}$ (Elastic-Net) which basically combines both of the previous ones. \cite[Section 4.1]{factor}

\subsection{Gradient Descent} \label{sec:gd}
We use $batch\ gradient\ descent$ (BGD) to figure out the values for $\theta$. 
An update to $\theta_j$ will therefore be computed by this formula:
$$\forall j\in \left\{ x \in \nat_0 \mid x \leq n\right\} : \theta_j := \theta_j - \alpha \cdot \frac{\partial}{\partial\theta_j}\Epsilon(\theta) \text{ \ \cite[p. 3]{learningLinear}\cite[p. 6]{factor}\cite{andrew_2}}$$
Where $\alpha$ is the learning rate that is used to determine the size of each step.
The derivative with respect to $\theta_j$ of the least squares objective function discussed in Section \ref{sec:obj} takes this form (for simplicity $y$ is also considered a feature with its corresponding $\theta$ fixed to -1):
$$ \frac{\partial}{\partial\theta_j}\Epsilon(\theta) = \sum_{i=1}^{m}h_\theta(x^{(i)}) \cdot x_j^{(i)} + \lambda \frac{\partial}{\partial\theta_j}R(\theta) \text{ \ \cite[p. 6]{factor}}$$

\subsection{Feature Scaling} \label{sec:featScal}
\begin{figure}[tb]
\centering
\begin{subfigure}[t]{0.49\linewidth}
\centering
\includegraphics[width=\linewidth]{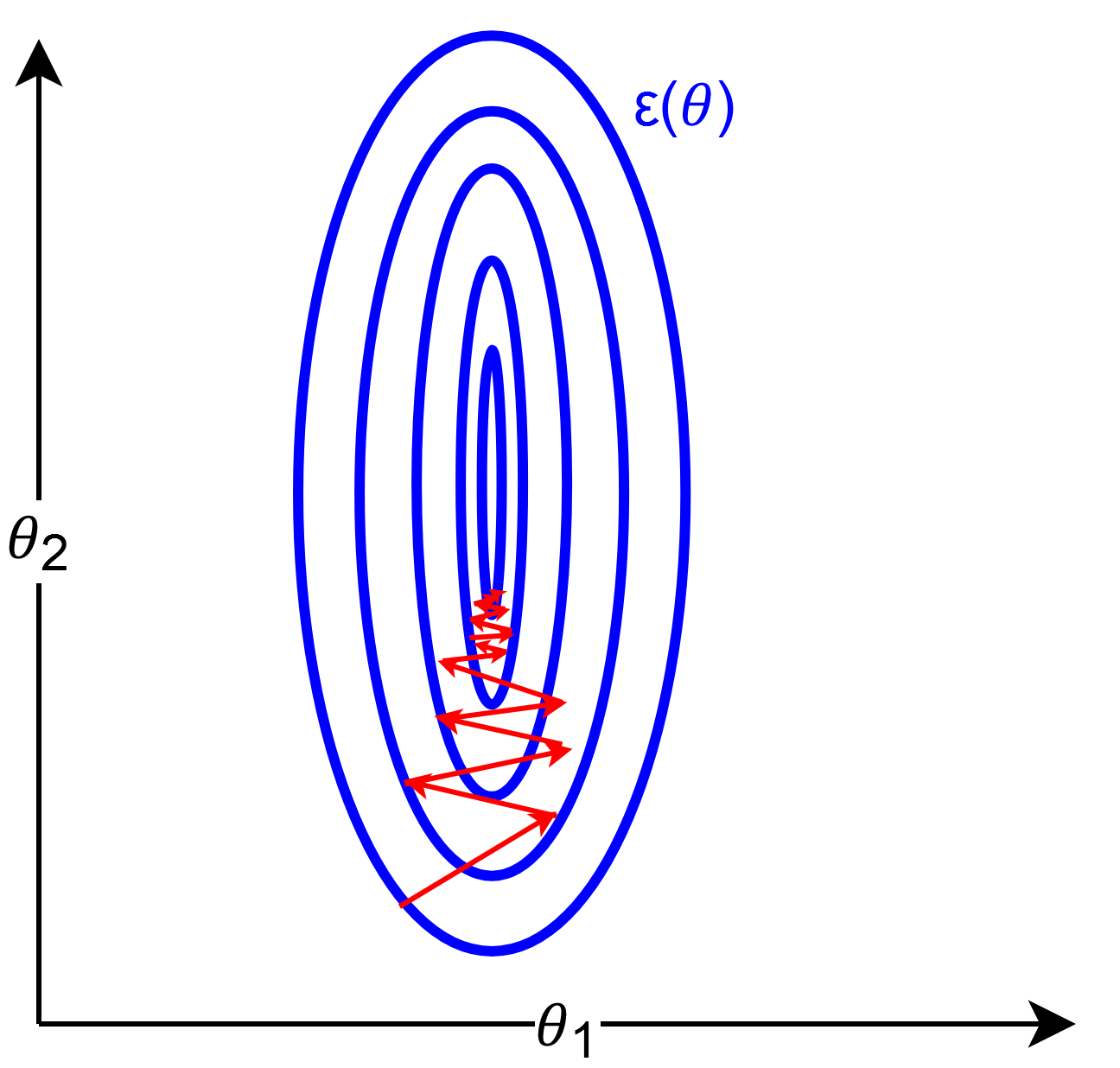}
\caption{Variables of different scales}
\label{fig:scaling_bad}
\end{subfigure}%
~
\begin{subfigure}[t]{0.49\linewidth}
\centering
\includegraphics[width=\linewidth]{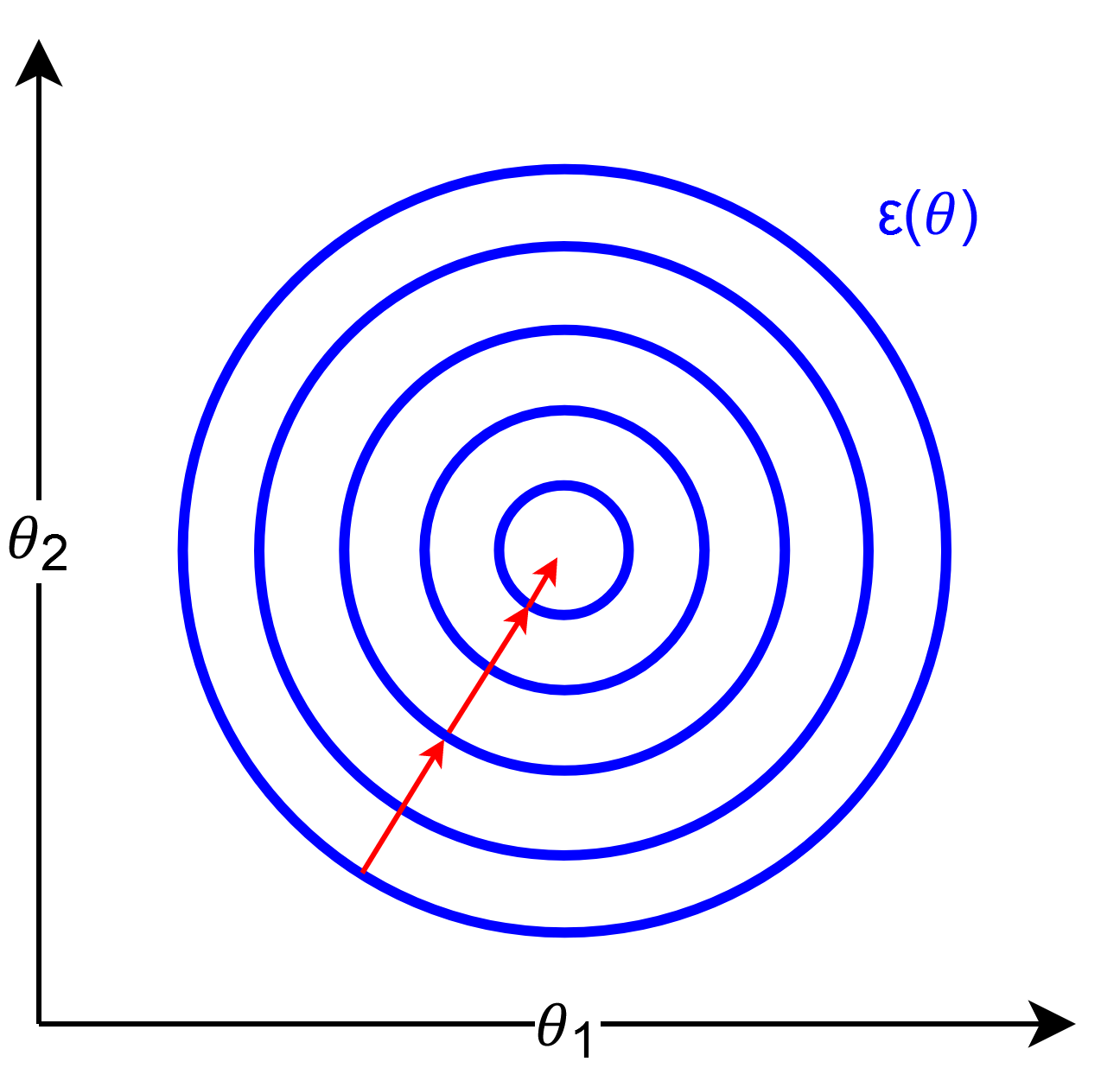}
\caption{Variables with similar scales}
\label{fig:scaling_good}
\end{subfigure}
\caption[Gradient descent with two variables of varying scales]{Gradient descent with two variables of varying scales}
\label{fig:scaling}
\end{figure}
\noindent
The idea of feature scaling (cf.~\autoref{fig:scaling}) is to scale all of the features involved before applying BGD.
Mean normalization consists of computing the average and maximum values over all $m$ rows for every feature $x_j$ and then computing new values such that all of them are in the interval $[-2,2] \ (\text{actually }[\frac{min-avg}{max}, \frac{max-avg}{max}])$:
$$avg_j = \frac{ \sum_{i=1}^m x_j^{(i)}} {m},\hspace{1cm} max_j = max\left\{\abs{x_j^{(i)}} \mid i\in [m]\right\}$$
$$\forall i \in [m]: x_{j, conv}^{(i)} := \frac{x_j^{(i)} - avg_j}{max_j} \text{ \ \cite{andrew_3}}$$
Note that since $\forall i \in [m]: x_0^{(i)}=1$ holds, feature $0$, which represents the constant offset, should never be rescaled.

After converting all values and executing gradient descent on the converted data, the computed values for $\theta_j$ have to be rescaled as well.
This way it is assured that they conform to the original range of values for $x_j$ instead of the rescaled range of $x_{j,conv}$:
{
\footnotesize
$$\sum_ {j=0}^n \theta_{j,conv} \cdot x_{j,conv} = \theta_{0,conv} \cdot 1 + \sum_ {j=1}^n \theta_{j,conv} \cdot \frac{x_j - avg_j}{max_j} \mbeq \sum_{j=0}^n \theta_j \cdot x_j$$
$$ \Leftrightarrow \ \theta_{0,conv}+ \sum_ {j=1}^n \theta_{j,conv} \cdot \left(\frac{x_j}{max_j} - \frac{avg_j}{max_j}\right) \mbeq \sum_{j=0}^n \theta_j \cdot x_j$$
$$ \Leftrightarrow \ \theta_{0,conv} + \sum_ {j=1}^n \theta_{j,conv} \cdot \frac{x_j}{max_j} - \sum_{j=1}^n \theta_{j,conv} \cdot \frac{avg_j}{max_j} \mbeq \sum_{j=0}^n \theta_j \cdot x_j$$
$$ \Leftrightarrow \ \left( \theta_{0,conv} - \sum_{j=1}\frac{\theta_{j,conv}}{max_j} \cdot avg_j \right) + \left( \sum_ {j=1}^n \frac{\theta_{j,conv}}{max_j} \cdot x_j \right) \mbeq \left( \theta_0 \cdot 1 \right)+ \left( \sum_{j=1}^n \theta_j \cdot x_j \right)$$
$$\Leftrightarrow \forall j \in [n]: \theta_{j} = \frac{\theta_{j,conv}} {max_j} \hspace{2mm} \text{and} $$
$$\theta_0 = \theta_{0,conv} - \sum_{j=1}^n \frac{\theta_{j,conv}}{max_j} \cdot avg_j = \theta_{0,conv} - \sum_{j=1}^n \theta_j \cdot avg_j$$
}
\begin{table}[tb]
\caption{Example data set with $m=5$ rows and $n=2$ features (not counting $x_0$)}
\label{tab:scaling}
\centering
\begin{subtable}{.5\linewidth}
\centering
\caption{Original values}
\label{tab:scaling_orig}
\small
\begin{tabular}{|c|c|c|c|c|}
\hline
$i$ & $y$ & $x_0$ & $x_1$ & $x_2$\\ \hline
1 & 2004 & 1 & 0.01 & 20000\\ \hline
2 & 5 & 1 & 0.03 & 0\\ \hline
3 &-1955 & 1 & -0.05 & -19500\\ \hline
4 & 999 & 1 & -0.01 & 10000\\ \hline
5 & -696 & 1 & 0.02 & -7000\\ \hline
\end{tabular}
\end{subtable}%
\begin{subtable}{.5\linewidth}
\caption{Rescaled values}
\label{tab:scaling_scaled}
\small
\centering
\begin{tabular}{|c|c|c|c|c|}
\hline
$i$ & $y$ & $x_0$ & $x_{1, conv}$ & $x_{2, conv}$\\ \hline
1 & 2004 & 1 & 0.2 & 0.965\\ \hline
2 & 5 & 1 & 0.6 & -0.035\\ \hline
3 & -1955 & 1 & -1 & -1.01\\ \hline
4 & 999 & 1 & -0.2 & 0.465\\ \hline
5 & -696 & 1 & 0.4 & -0.385\\ \hline
\end{tabular}
\end{subtable}
\end{table}
\noindent

\textbf{Example}~(cf.~Table \ref{tab:scaling_orig}):
First we compute the average and maximum values for $j \in [2]$:
$$avg_1 = \frac{ \sum_{i=1}^5 x_1^{(i)}} {5} = \frac{0.01+0.03-0.05-0.01+0.02}{5} = 0$$
$$ max_1 = max\left\{\abs{x_1^{(i)}} \mid i\in [5]\right\} = max\left\{0.01,0.03,0.05,0.02\right\} = 0.05$$
$$avg_2 = \frac{ \sum_{i=1}^5 x_2^{(i)}} {5} = \frac{20000+0-19500+10000-7000}{5} = 700$$
$$ max_2 = max\left\{\abs{x_2^{(i)}} \mid i\in [5]\right\} = 20000$$
Using these results to rescale the original data for $x_1$ and $x_2$ leads us to the values shown in Table \ref{tab:scaling_scaled}.
After applying BGD to those rescaled values, we obtain the following results (rounded, depends on accuracy set for BGD) for $\theta$ which we will rescale as well:
$$\theta_{1, conv} = 10 \Rightarrow \theta_{1} = \frac{\theta_{1,conv}}{max_1} = \frac{10}{0.05} = 200$$
$$\theta_{2, conv} = 2000 \Rightarrow \theta_{2} = \frac{\theta_{2,conv}}{max_2} =\frac{ 2000}{20000} = 0.1$$
$$\theta_{0,conv} = 70 \Rightarrow \theta_{0} = \theta_{0,conv} - (\theta_{1}\cdot avg_1 + \theta_{2}\cdot avg_2) = 0$$
Finally, we insert all of the values for $\theta$ into $h_\theta(x)$ to obtain the following function:
$$h_\theta(x) = 200 \cdot x_1 + 0.1\cdot x_2$$

\subsection{Separating Aggregates and Convergence of Gradient Descent} \label{sec:rewriting}
Normally, computing the value of $\frac{\partial}{\partial\theta_j}\Epsilon(\theta)$ requires scanning the complete data in every iteration which results in a significant overhead. However, by rewriting the function it is possible to precompute the data dependent part, leaving only less intensive operations for the convergence that will have to be repeated several times.\\
To transform it we only need to look at the part of $\frac{\partial}{\partial\theta_j}\Epsilon(\theta)$ without the regularization term which we will call $S_j$:
$$ S_j = \sum_{i=1}^{m}h_\theta(x^{(i)}) \cdot x_j^{(i)} = \sum_{i=1}^{m}\sum_{k=0}^n (\theta_k \cdot x_k^{(i)}) \cdot x_j^{(i)}
= \sum_{i=1}^{m}\sum_{k=0}^n (\theta_k \cdot x_k^{(i)}) \cdot x_j^{(i)} $$
$$= \sum_{k=0}^n \sum_{i=1}^{m} (\theta_k \cdot x_k^{(i)}) \cdot x_j^{(i)} = \sum_{k=0}^n \theta_k \cdot \sum_{i=1}^{m} x_k^{(i)} \cdot x_j^{(i)}
= \sum_{k=0}^n \theta_k \cdot Cofactor[k,j]$$
$$\text{with } Cofactor[k,j] = \sum_{i=1}^{m} x_k^{(i)} \cdot x_j^{(i)}$$
This allows us to precompute all of the cofactors to obtain a cofactor matrix, which can then be used to update $\theta$ during gradient descent without being dependent on the data.
This matrix has the following properties, given a database $D$ and a query $Q$ with the schema of the query result $Q(D)$ being $\sigma = (A_l)_{l \in [n]}$:
\begin{itemize}
\item symmetry: $\forall k,j \in [n]: Cofactor[k,j]=Cofactor[j,k]$ \cite[Proposition 4.1]{factor} \cite[Proposition 3.1]{learningLinear}
\item commutativity with union: Given a disjoint partitioning\\$D=\bigcup_{j\in [p]}(D_j)$ and cofactors $(Cofactor_j)_{j\in [p]}$ over \\ $(Q(D_j))_{j\in [p]}$ with database $D$ and query $Q$:\\
$\forall k,i \in [n]: Cofactor[k,i] = \sum_{j=1}^p Cofactor_j[k,i]$ \cite[Proposition 4.1]{factor} \cite[Proposition 3.1]{learningLinear}
\item commutativity with projection: feature set $L\subseteq \sigma$ and cofactor matrix $Cofactor_L$ for the training dataset $\pi_L(Q(D))$:\\
$\forall k,j \in [n]: A_k,A_j \in L \rightarrow Cofactor_L[k,j] = Cofactor[k,j]$ \cite[Proposition 4.1]{factor} \cite[Proposition 3.1]{learningLinear}
\end{itemize}
Due to the symmetry of $Cofactor$, only about half of the matrix has to be computed: $\left(\frac{n\cdot(n+1)}{2} \text{ instead of } n^2 \right)$.
The commutativity with union allows computing cofactors over multiple training data sets and then combining the results by summing up the cofactor matrices \cite[p. 7]{factor}.
Commutativity with projection enables us to simply ignore parameters not required for the regression, which may however be needed for the join computation \cite[p. 7]{factor}.

\section{Implementation} \label{sec:impl}
This section describes the implementation of linear regression in C++ using cofactors and BGD.
It uses pqxx \cite{pqxx} to connect to a PostgreSQL \cite{psql} database that contains the training data.
Most operations discussed in Section \ref{sec:fact} and \ref{sec:linReg} will be computed by SQL queries.
\autoref{fig:decl} shows the declarations of the functions discussed in this section.
\begin{lstlisting}[frame=single, language=C++,
,caption={Declarations of functions described in §\ref{sec:scaling}-\ref{sec:combining}.}
,label={fig:decl}
,float
]
vector<scaleFactors> scaleFeatures(const vector<string>& relevantColumns, vector<ExtendedVariableOrder*>& leaves, const string& con);
void factorizeSQL(const ExtendedVariableOrder& varOrder, const string& con);
vector<double> batchGradientDescent(const vector<string>& relevantColumns, pqxx::connection& con);
vector<double> linearRegression(ExtendedVariableOrder& varOrder, const vector<string>& relevantColumns, const string& con, double& avg);
\end{lstlisting}

\subsection{Variable Order} \label{sec:varOrder}
\begin{figure}[tb]
\centering
\includegraphics[width=\linewidth]{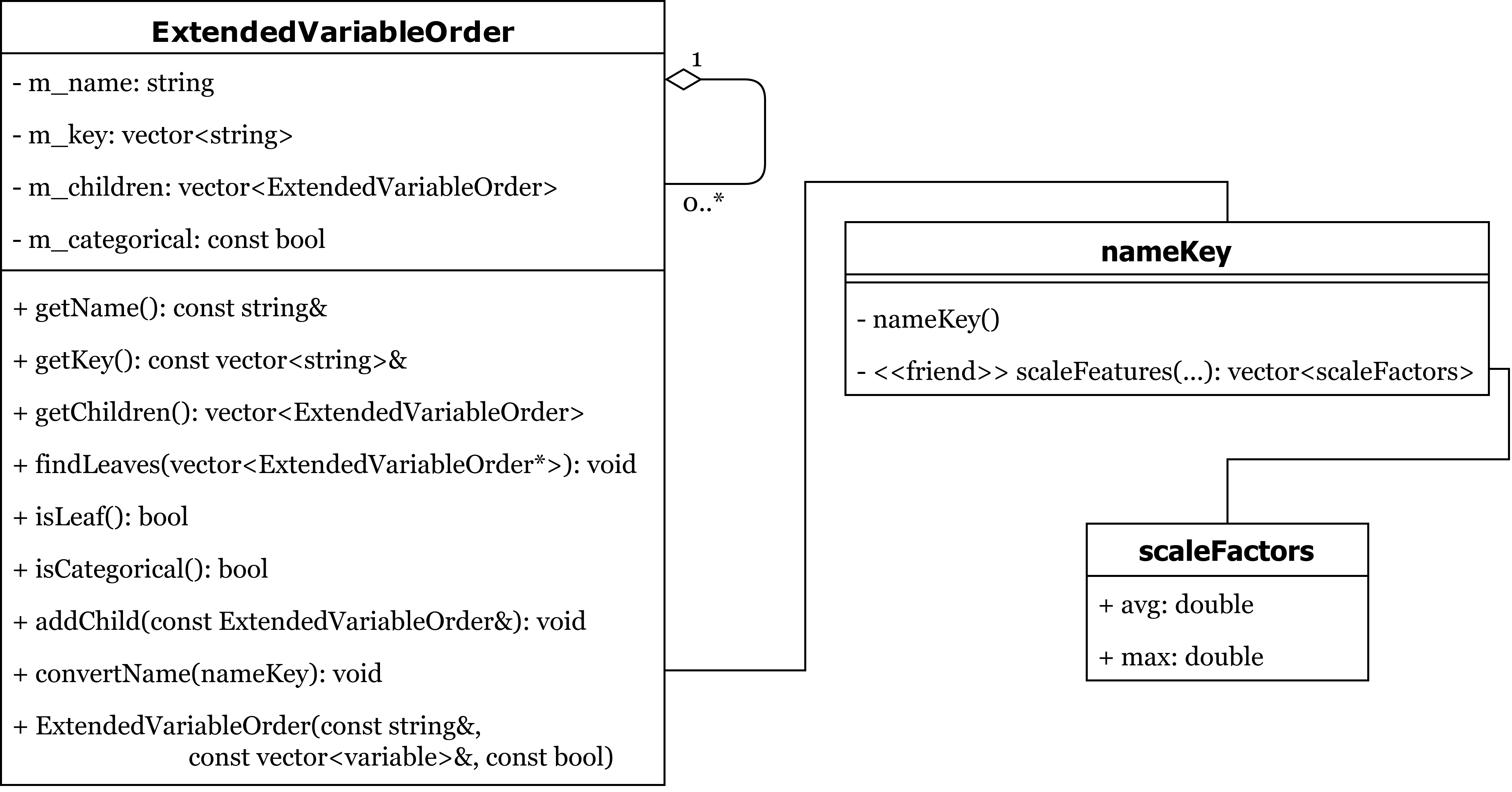}
\caption[UML diagram showing the structure of the class $ExtendedVariableOrder$, its inner class $nameKey$ and the struct $scaleFactors$]{UML diagram showing the structure of the class $ExtendedVariableOrder$, its inner class $nameKey$ and the struct $scaleFactors$}
\label{fig:uml}
\end{figure}

\begin{figure}[tb]
\centering
\includegraphics[width=\linewidth]{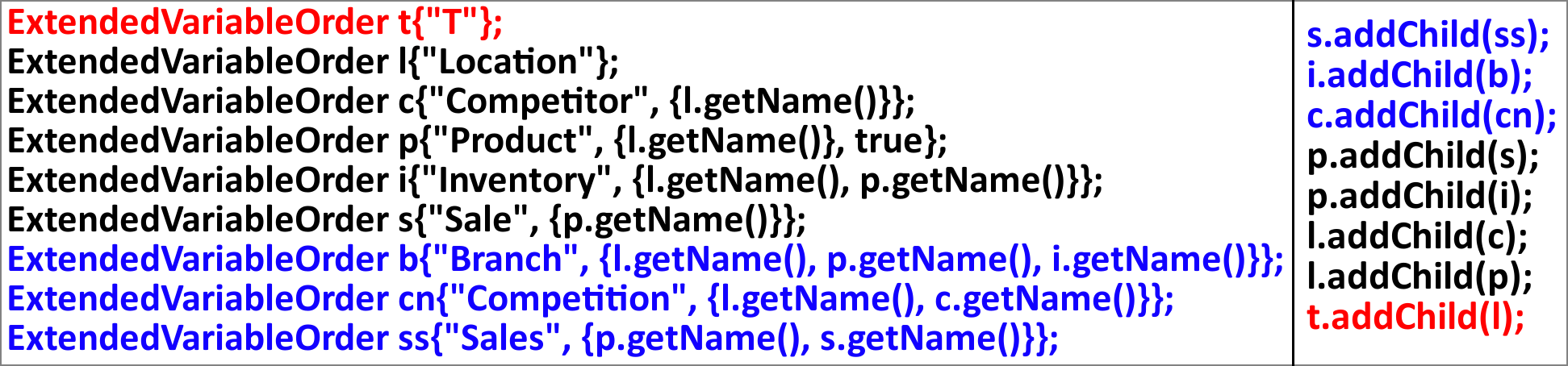}
\caption[Construction of an extended variable order for the schema given in \autoref{fig:sale}]{Construction of an extended variable order for the schema given in \autoref{fig:sale}. Lines interacting with the {\color{red}intercept} are marked in {\color{red}red} while lines referring to {\color{blue}relations} are highlighted in {\color{blue}blue}.}
\label{fig:evo}
\end{figure}

The class $ExtendedVariableOrder$ (cf.~\autoref{fig:uml}) represents a variable order as described in Section \ref{sec:vaOr}. 
It is called $extended$ because a complete variable order is assumed to fulfill these additional properties:
\begin{enumerate}
\item The lowest (greatest distance to the root) variable of a relation $R$ in a variable order $\Delta$ has a child representing the relation $R$. This additional node stores all of its attributes in its corresponding $key(R)$.
Since all attributes in the same relation appear within a simple path from the root, the variable to which $R$ is appended is unique. Furthermore, every leaf node in some variable order $\Delta$ is part of at least one relation and therefore followed by one or multiple new nodes.
This leads to the following property: $\forall q: isLeaf(q) \Leftrightarrow q$~represents a relation.
\item The intercept will be represented by an additional node $T$ in the extended variable order. It will be the new root of the variable order and has all of the roots of the original variable order as its children.
\end{enumerate}
\if false
These are the functions of the class $ExtendedVariableOrder$:
\begin{itemize}
\item $getName()$ returns the name of this variable which is stored in $m\_name$.
\item $getKey()$ returns all the names of the variables it depends on which are stored in $m\_key$.
\item $getChildren()$ returns all the children of this variable which are stored in $m\_children$.
\item $findLeaves(vector\textless ExtendedVariableOrder\textgreater)$ adds all of the leaves of the subtree rooted at this node to the vector which it is given a reference to.
\item $isLeaf()$ returns true if this variable does not have any children and is therefore a leaf; otherwise it returns false.
\item $isCategorical()$ returns whether this variable is categorical or not which is indicated by $m\_categorical$. An attribute is categorical, if it's not represented by numerical values. The attribute $city$ might for example be categorical and have the possible values $\{ Munich, New\ York, Berlin, Washington \}$
\item $addChild(const\ ExtendedVariableOrder\&)$ adds the variable order given as a parameter, as a subtree of this node by updating $m\_children$.
\item $convertName(nameKey)$ appends $\_conv$ to the name of this variable to signal it has been converted in feature scaling. Since this method should only be accessible from the $scaleFeatures(\dots)$ function described in Section \ref{sec:scaling}, it requires a parameter of type $nameKey$. This $nameKey$ only has a private constructor and can therefore only be created from $scaleFeatures(\dots)$ because it is the only function marked as a $friend$ of this class. This access-protecting pattern is called the $passkey$ pattern \cite{passkey}.
\item $ExtendedVariableOrder(const\ string\&, const\ vector\textless variable\textgreater\&, const\ bool)$ is the constructor of this class. Its parameters are: A string which will be used as the name for this variable ($m\_name$), a list of all of the variables it depends on ($m\_key$) and a bool which indicates whether the variable is categorical or not ($m\_categorical$). This last parameter is optional and defaults to false. It also initializes $m\_children$ with an empty vector.
\end{itemize}
\fi
\autoref{fig:evo} shows how the extended variable order corresponding to the variable order shown in \autoref{fig:sale_vo} can be constructed if the attribute $Product$ is categorical and all other columns are of numerical value.

\subsection{Feature Scaling on a Variable Order} \label{sec:scaling}
The function $scaleFeatures(\dots)$ takes as input a vector of strings $relevantColumns$, a vector of pointers to ExtendedVariableOrder nodes $leaves$ and a string $con$ and returns a vector of $scaleFactors$ consisting of an average and a maximum value.
\if false
\begin{itemize}
\item $relevantColumns$ consists of entries $(col_0, col_1, \dots, col_n)$ which should contain all the names of features relevant for learning in $col_1$ to $col_{n-1}$. Additionally, $col_0$ should be the name of the label we want to compute (not necessary for all the versions discussed in Section \ref{sec:favorita} but some require it). The last entry $col_n$ will simply be ignored and exists only for the purpose of allowing the reuse of $relevantColumns$ in other functions.
\item $leaves$ is supposed to contain all of the leaves of the extended variable order$\Delta$ or at least all of the leaves containing some feature or label. One way to create such a vector is by calling $findLeaves$ on the root/intercept of $\Delta$.
\item $con$ is a string containing all necessary information to connect to the database containing the relations from $leaves$ via a $pqxx$::$connection(con)$.
\item $scaleFactors$ will have $n$ entries $(agg_0,agg_1, \dots, agg_{n-1})$ where $agg_i$ represents the average and maximum value of $col_i$ over all relations it appears in.
\end{itemize}
\fi

First of all it is identified, which $col_i$ occurs in which leaves.
In the second step the $scaleFactors$ are computed for all the features by running a SQL query taking the union of all tables containing $col_i$ and computing the average and maximum (absolute values) over this union.
This ensures to scale every occurrence of a feature by the same factors even if there might be different values in different relations.
Therefore, equijoins will still work even after rescaling since $x=y \Leftrightarrow \frac{x-a}{b} = \frac{y-a}{b}$.
Running those queries is parallelised for the different features using OpenMP \cite{openmp}.
This requires a new $pqxx$::$connection$ for every attribute $col_i$.
As a last step, for every relation that has at least one rescaled attribute, a new view will be created.
Afterwards, the function terminates and returns the $scaleFactors$ vector.
An execution of this function on the example schema from \autoref{fig:sale} with the extended variable order created by \autoref{fig:evo} results in the SQL queries shown in \autoref{fig:scaling_sql}.
\begin{lstlisting}[frame=single, language=SQL,mathescape=true,escapeinside=\`\`
,float
,caption={SQL queries for an input corresponding to the variable order created by \autoref{fig:evo}:\newline
$relevantColumns=("Inventory", "Competitor", "Sale", "T")$;\newline$leaves=("Competition'',"Sales","Inventory")$\newline
The values highlighted in {\color{red}red} are the result values of the first two queries and depend on the data}
,label={fig:scaling_sql}
]
WITH unionOfAllTables AS (SELECT Competitor FROM Competition) 
    SELECT AVG(COALESCE(Competitor,0)) as avg, MAX(ABS(COALESCE(Competitor,0))) AS max 
    FROM unionOfAllTables;
WITH unionOfAllTables AS (SELECT Sale FROM Sales) 
    SELECT AVG(COALESCE(Sale,0)) as avg,  MAX(ABS(COALESCE(Sale,0))) AS max 
    FROM unionOfAllTables;

CREATE VIEW Sales_conv AS 
    SELECT Product, ((Sale - `\color{red}-11.526316`) / `\color{red}188.000000`)::real AS Sale 
    FROM Sales;
CREATE VIEW Competition_conv AS 
    SELECT Location, ((Competitor - `\color{red}-22.517241`) / `\color{red}174.000000`)::real AS Competitor 
    FROM Competition;
\end{lstlisting}

\subsection{Computing Cofactors over a Factorized Join} \label{sec:fac}
The function $factorizeSQL(\dots)$ computes the aggregates required for the cofactor matrix discussed in Section \ref{sec:rewriting}.
It has two parameters:
\begin{itemize}
\item $varOrder$ is the root/intercept of an $ExtendedVariableOrder$ of a factorized representation of some join as explained in Section~\ref{sec:varOrder}.
\item $con$ is a string containing all necessary information to connect to the database containing the relations occurring in $varOder$ using a $pqxx$::$connection(con)$.
\end{itemize}
At first, it establishes a $pqxx$::$connection$ $con$ to the database specified in $con$.
Then it iterates over all of its children $x$ to call a helper function with subtree $x$ and the connection $con$ as arguments.
After all children have been processed, it will run a query to create its own table named $"Q"+varOrder.getName()$, which will combine all of the intermediate results to a final result.

The helper function \texttt{factorizeSQL(ExtendedVariableOrder\ x, pqxx$::$connection\ con)} distinguishes between two cases and is based on the pseudo code described in \textit{Factorized Databases}~\cite[Figure 5]{factor}.
For better readability the name of the variable $x$ will be called $xName$.

\subsubsection{Leaf Nodes}
Nodes without children will first create a table named $xName+"\_type"$ which contains a single row with two columns: The name of this relation $(xName+"\_n")$ and the value 0 $(xName+"\_d")$, since leaf nodes represent only constants and no linear or quadratic terms.
Then it will create a new view named $"Q"+xName$.
It contains all the columns in its schema, which is given by $x.getKey()$, a lineage column $xName+"\_lineage"$ containing an empty string, a column representing $xName+"\_d"$ called $xName+"\_deg"$ and a column $xName+"\_agg"$ with value $1$.

\subsubsection{Inner Nodes}
Variables with children will also create a table $xName+"\_type"$ first but it contains three rows in total: One for each value in $\{0,1,2\}$ as $(xName+"\_d")$ combined with $(xName+"\_n")$ which contains $xName$.
Then it will iterate over its children and prepare its own query for its own view $"Q"+xName$:
\begin{itemize}
\item The variables it depends on will still be represented in the new view by selecting them from the children in the same relation as this one.
\item The column $xName+"\_agg"$ will be filled by a sum aggregate over the multiplication of $x^{xName\_d}$ and the $agg$ values of its children. If $x$ is categorical $x^{xName\_d}$ will be replaced by the value 1.
\item $xName+"\_deg"$ will be defined as the sum of the $deg$ values of $x$'s children and $xName\_d$.
\item Its lineage will consist of the concatenation of $x$'s children's lineages and will contain $xName\_n$ only if $xName\_d$ is greater than 0.
\end{itemize}
All of those values will be determined over the join of its subviews and $xName\_type$.
Additionally, they will be filtered so that only those with $xName\_deg \leq 2$ will be kept.

\subsubsection{Example}
\autoref{fig:fact_table} shows the SQL statements used for the creation of the $x\_type$ tables for the path corresponding to the $Competition$ relation.
The variable order, which this subquery is taken from, is the one created in \autoref{fig:evo} after the data has been rescaled by $scaleFeatures(\dots)$ from Section \ref{sec:scaling}.
\begin{lstlisting}[frame=single, language=SQL
,float
,caption={SQL statements created by factorizeSQL for the relation $Competition$}
,label={fig:fact_table}
]
CREATE TABLE Location_type(Location_n varchar(50),Location_d int);
INSERT INTO Location_type VALUES('Location', 0);
INSERT INTO Location_type VALUES('Location', 1);
INSERT INTO Location_type VALUES('Location', 2);

CREATE TABLE Competitor_type(Competitor_n varchar(50), Competitor_d int);
INSERT INTO Competitor_type VALUES('Competitor', 0);
INSERT INTO Competitor_type VALUES('Competitor', 1);
INSERT INTO Competitor_type VALUES('Competitor', 2);

CREATE TABLE Competition_conv_type(Competition_conv_n varchar(50),
                                   Competition_conv_d int);
INSERT INTO Competition_conv_type VALUES ('Competition_conv', 0);
\end{lstlisting}

\begin{figure}[htb]
\centering
\begin{subfigure}[t]{1\linewidth}
\begin{tikzpicture}[every node/.append style={inner sep=0pt, font=\vphantom{$f_p$}},
                    every path/.append style={shorten >=2.5pt, shorten <=2.5pt}
                   ,node distance=11pt]
\node(s1) {$\pi$};
\node [right=1pt of s1] (e1) {$_{lineage,\ deg,\ agg}$ = Table QT};
\node [below=of s1] (s2) {$\Gamma$};
\node [right=1pt of s2] (e2) {$_{lineage,\ L\_deg}$};
\node [below=of s2] (s3) {$\rho$};
\node [right=1pt of s3] (e3) {$_{lineage \leftarrow L\_lineage,\ deg\leftarrow L\_deg,\ agg \leftarrow SUM(L\_agg)}$};
\node [below=of s3] (s4) {$\sigma$};
\node [right=1pt of s4] (e4) {$_{L\_deg\ \leq\ 2}$};

\node [below=of s4] (s5) {$\pi$};
\node [right=1pt of s5] (e5) {$_{L\_lineage,\ L\_deg,\ L\_agg}$ = View QL};
\node [below=of s5] (s6) {$\Gamma$};
\node [right=1pt of s6] (e6) {$_{L\_lineage,\ C\_deg + P\_deg + L\_d}$};
\node [below=of s6] (s7) {$\rho$};
\node [right=1pt of s7,align=left] (e7) {\\$_{L\_lineage \leftarrow C\_lineage||P\_lineage||\dots,}$\\$_{deg\leftarrow C\_deg + P\_deg + L\_d,\ agg \leftarrow SUM(POWER(\dots)*C\_agg*P\_agg)}$};
\node [below=of s7] (s8) {$\sigma$};
\node [right=1pt of s8] (e8) {$_{L\_lineage,\ C\_deg + P\_deg + L\_d\ \leq\ 2}$};
\node [below=of s8] (s9) {$\times$};
\node [below=of s9] (s11) {L\_type};
\node [right=of s11] (s10) {$\bowtie$};
\node [right=5pt of s10] (e10) {$_{QC.L=QP.L}$};
\node [below=of s10] (s12) {\dots = View QC};
\node [right=of s12] (s13) {\dots = View QP};

\draw (s1) -- (s2);
\draw (s2) -- (s3);
\draw (s3) -- (s4);

\draw (s4) -- (s5);

\draw (s5) -- (s6);
\draw (s6) -- (s7);
\draw (s7) -- (s8);
\draw (s8) -- (s9);
\draw (s9) -- (s10);
\draw (s9) -- (s11);
\draw (s10) -- (s12);
\draw (s10) -- (s13);
\end{tikzpicture}
\caption{Operator tree for the top of $QT$ and $QL$}
\end{subfigure}
\begin{subfigure}[b]{1\linewidth}
\centering
\begin{tikzpicture}[every node/.append style={inner sep=0pt, font=\vphantom{$f_p$}},
                    every path/.append style={shorten >=2.5pt, shorten <=2.5pt}
                   ,node distance=10pt]
\node(s1) {$\pi$};
\node [right=1pt of s1] (e1) {$_{L,\ C,\ Com\_conv\_lineage,\ Com\_conv\_deg}$ = View QCom\_conv};
\node [below=of s1] (s2) {$\rho$};
\node [right=1pt of s2] (e2) {$_{Com\_conv\_lineage \leftarrow '',\ Com\_conv\_deg \leftarrow Com\_conv\_d,\ Com\_conv\_agg \leftarrow 1}$};
\node [below=of s2] (s3) {$\times$};
\node [below=of s3] (s4) {Com\_conv};
\node [right=15pt of s4] (s5) {Com\_conv\_type};

\node [above=of s1] (d1) {$\dots$};
\draw (d1) -- (s1);

\draw (s1) -- (s2);
\draw (s2) -- (s3);
\draw (s3) -- (s4);
\draw (s3) -- (s5);
\end{tikzpicture}
\caption{Operator tree for $QCom\_conv$}
\end{subfigure}
\caption{Part of the operator tree of $QT$ corresponding to $QT$, $QL$ and $QCom\_conv$}
\label{fig:optree}
\end{figure}
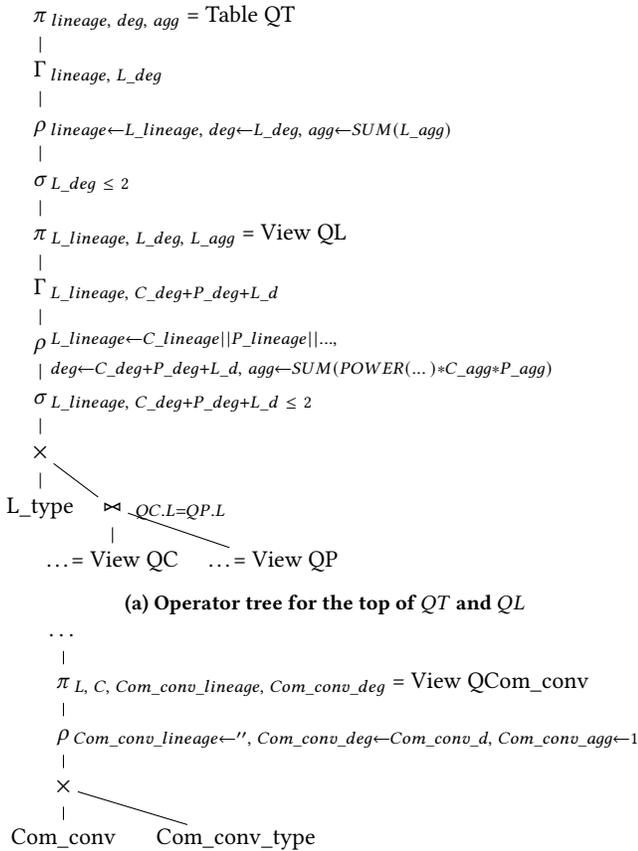

\autoref{fig:fact_view} shows all of the SQL queries that are used to create the $QxName$ views as well as the query for the intercept.
The variable order is again the same as created in Section \ref{sec:varOrder} after it was processed by $scaleFeatures(\dots)$.
\autoref{fig:optree} shows the parts of the operator tree of $QT$ that corresponds to the queries of the root $QT$, the inner node $QL$ and the leaf node $QCom\_conv$.

\begin{lstlisting}[frame=single,language=SQL
,float
,caption={SQL queries run by $factorizeSQL(\dots)$ for all of the variables}
,label={fig:fact_view}
]
CREATE VIEW QCom_conv AS (
SELECT Com_conv.L, Com_conv.C, ''::text AS Com_conv_lineage, Com_conv_d AS Com_conv_deg, 1 AS Com_conv_agg 
FROM Com_conv, Com_conv_type);

CREATE VIEW QC AS (
SELECT QCom_conv.L, 
       QCom_conv.Com_conv_lineage || 
        CASE WHEN C_d > 0 THEN '(' || C_n || ',' || C_d || ')' ELSE ''::text END AS C_lineage, 
       QCom_conv.Com_conv_deg + C_d AS C_deg, 
       SUM(POWER(COALESCE(QCom_conv.C,0),C_d) * QCom_conv.Com_conv_agg) AS C_agg 
FROM QCom_conv, C_type 
WHERE QCom_conv.Com_conv_deg + C_d <= 2 
GROUP BY QCom_conv.L, C_lineage, QCom_conv.Com_conv_deg + C_d);

CREATE VIEW QSa_conv AS (
SELECT Sa_conv.P, Sa_conv.S, 
       ''::text AS Sa_conv_lineage, Sa_conv_d AS Sa_conv_deg, 
       1 AS Sa_conv_agg 
FROM Sa_conv, Sa_conv_type);

CREATE VIEW QS AS (
SELECT QSa_conv.P, 
       QSa_conv.Sa_conv_lineage || 
        CASE WHEN S_d > 0 THEN '(' || S_n || ',' || S_d || ')' ELSE ''::text END AS S_lineage, 
       QSa_conv.Sa_conv_deg + S_d AS S_deg, 
       SUM(POWER(COALESCE(QSa_conv.S,0),S_d) * QSa_conv.Sa_conv_agg) AS S_agg 
FROM QSa_conv, S_type 
WHERE QSa_conv.Sa_conv_deg + S_d <= 2 
GROUP BY QSa_conv.P, S_lineage, QSa_conv.Sa_conv_deg + S_d);

CREATE VIEW QBr AS (
SELECT Br.L, Br.P, Br.I, 
       ''::text AS Br_lineage, Br_d AS Br_deg, 1 AS Br_agg 
FROM Br, Br_type);

CREATE VIEW QI AS (
SELECT QBr.P, QBr.L, 
       QBr.Br_lineage || 
        CASE WHEN I_d > 0 THEN '(' || I_n || ',' || I_d || ')' ELSE ''::text END AS I_lineage, 
       QBr.Br_deg + I_d AS I_deg, 
       SUM(POWER(COALESCE(QBr.I,0),I_d) * QBr.Br_agg) AS I_agg 
FROM QBr, I_type 
WHERE QBr.Br_deg + I_d <= 2 
GROUP BY QBr.P, QBr.L, I_lineage, QBr.Br_deg + I_d);

CREATE VIEW QP AS (
SELECT QI.L, 
       QS.S_lineage || QI.I_lineage || 
        CASE WHEN P_d > 0 THEN '(' || P_n || ',' || P_d || ')' ELSE ''::text END AS P_lineage, 
       QS.S_deg + QI.I_deg + P_d AS P_deg, 
       SUM(1 * QS.S_agg * QI.I_agg) AS P_agg 
FROM QS JOIN QI ON QS.P=QI.P, P_type 
WHERE QS.S_deg + QI.I_deg + P_d <= 2 
GROUP BY QS.P, QI.L, P_lineage, QS.S_deg + QI.I_deg + P_d);

CREATE VIEW QL AS (
SELECT QC.C_lineage || QP.P_lineage || 
        CASE WHEN L_d > 0 THEN '(' || L_n || ',' || L_d || ')' ELSE ''::text END AS L_lineage, 
       QC.C_deg + QP.P_deg + L_d AS L_deg, 
       SUM(POWER(COALESCE(QC.L,0),L_d) * QC.C_agg * QP.P_agg) AS L_agg 
FROM QC JOIN QP ON QC.L=QP.L, L_type 
WHERE QC.C_deg + QP.P_deg + L_d <= 2 
GROUP BY L_lineage, QC.C_deg + QP.P_deg + L_d);

CREATE TABLE QT AS (
SELECT QL.L_lineage AS lineage, 
       QL.L_deg AS deg, 
       SUM(QL.L_agg) AS agg 
FROM QL 
WHERE QL.L_deg <= 2 
GROUP BY lineage, QL.L_deg);
\end{lstlisting}

\subsection{Batch Gradient Descent using Cofactors} \label{sec:bgd}
The function $batchGradientDescent(\dots)$ performs gradient descent on a precomputed cofactor matrix that is stored in a database.
At first, this function constructs a cofactor matrix locally by requesting the data from the database.
To identify which cofactor represents which feature, the lineage and deg columns are used.
For example $cofactor[0][1] = cofactor[1][0]$ would be requested by the query:
\begin{lstlisting}[frame=single, language=SQL,mathescape=true]
SELECT $agg$
FROM $Qcol_n$
WHERE lineage LIKE '%($col_0$,1)%' AND lineage LIKE '%($col_1$,1)%';
\end{lstlisting}
And $cofactor[i][n] = cofactor[n][i]$ would be requested by this query:
\begin{lstlisting}[frame=single, language=SQL,mathescape=true,escapeinside=\`\`]
SELECT $agg$
FROM $Qcol_n$
WHERE lineage LIKE '%($col_i$,1)%' AND deg = 1;
\end{lstlisting}

After this matrix is constructed, $\theta$ will be initialized with $n+1$ values with $\theta_0=-1$.
Then it will repeat updating $\theta$ until the new change to the error is smaller than a threshold $\epsilon$ which is currently set to $10^{-6}$.
It will also abort if it takes more than 100 million iterations or if $\alpha$ is decreased below $10^{-15}$.
Initially it will start with $\alpha = 0.003$ and divide it by 3 every time the sum of all $\epsilon_j$ is greater than it was in the previous iteration.
The update for $\theta_j$ is computed as discussed in Section \ref{sec:gd} and \ref{sec:rewriting} by this formula:
$$\epsilon_j = \alpha \cdot \left( \sum_{k=0}^n \left( \theta_k \cdot Cofactor[k,j] \right) + 0.006 \cdot \theta_j \right)$$
Where $0.006 \cdot \theta_j$ replaces $\lambda \frac{\partial}{\partial\theta_j}R(\theta)$ for the ridge regularization term used in this implementation.

\subsection{Combining Cofactors and Linear Regression} \label{sec:combining}
The function $linearRegression$ combines all of the previous functions into one.
It calls $varOrder.findLeaves(leaves)$ to construct a vector with all the relations in the given extended variable order.
To convert the data, this vector will then be used to call the function $scaleFeatures(relevantColumns, leaves, con)$.
Afterwards, it will compute the cofactors required for gradient descent by calling $factorizeSQL(varOrder, con)$.
Once this is completed, 
a call to $batchGradientDescent()$ returns $\theta$.
However those values still correspond to the rescaled features, so $\theta_1$ to $\theta_{n-1}$ are rescaled as discussed in Section \ref{sec:featScal} by dividing $\theta_j$ by $scaleAggs_j.max$.
In the same iteration the sum of $\theta_j \cdot scaleAgg_j.avg$ is computed and used to update $\theta_n$ by subtracting this sum.
At last the function returns the adjusted $\theta$ and terminates.

\section{Evaluation} \label{sec:Eval}
This section compares non-factorized learning to different implementations for factorized learning with regard to runtime and accuracy.
We further benchmark our implementation on PostgreSQL and HyPer.

\if false
\subsection{Different Feature Scaling Mechanisms}\label{sec:diffFeatScaling}
Section \ref{sec:featScal} used mean normalization to rescale the features.
This section will cover some other ways of rescaling and provide a comparison between them on some randomly generated data sets.
Every data set follows the schema shown in \autoref{fig:sale}.
The error values always correspond to a run on 1000 different random generated data sets for each version.
Only versions with the same number as prefix (e.g 1.1 and 1.2) use the same data.

Each data set has 9 to 10000 rows and is set up so that it can be processed by the variable order constructed in \autoref{fig:evo}. Furthermore, the data is created such that it corresponds to some randomly generated function $h_{\theta_{expected}}(x)$. This allows easy computation of the error of the result returned by the linear regression. Those values $\theta_{expected}$ are also created randomly within the range $[-200.0,200.0]$.
Table \ref{tab:scaling_rel} provides the relative errors for the values of $\theta$ for the features which are calculated by this formula: $$relError=\abs{\frac{\theta_j - \theta_{j,expected}}{\theta_{j,expected}}}$$
\begin{table}[tb]
\centering
\begin{tabular}{|c|c|c|c|} \hline
\textbf{version} & \textbf{max. rel. error} & \textbf{min. rel. error} & \textbf{avg. rel. error} \\ \hline
1.1 & 0.0558168 \% & 0.000156102 \% & 0.00619048 \% \\
1.2 & 0.1667480 \% & 0.000621344 \% & 0.01873150 \% \\ \hline
2.1 & 3.3960700 \% & 0.000159206 \% & 0.00986404 \% \\
2.2 & 3.7434900 \% & 0.000159049 \% & 0.00995783 \% \\ \hline
3.1 & 0.3815420 \% & 0.000163301 \% & 0.00629162 \% \\
3.2 & 0.3815420 \% & 0.000163301 \% & 0.00629162 \% \\ \hline
\end{tabular}
\caption[Maximum, minimum and average relative errors for $\theta_j$ of different scaling versions over 1000 randomly generated data sets]{Maximum, minimum and average relative errors for $\theta_j$ of different scaling versions over 1000 randomly generated data sets. Only versions with the same prefix are run on the exact same data and directly comparable.}
\label{tab:scaling_rel}
\begin{tabular}{|c|c|c|c|} \hline
\textbf{version} & \textbf{max. rel. error} & \textbf{min. rel. error} & \textbf{avg. rel. error} \\ \hline
1.1 & 441.532 & 0.00218998 & 16.8114 \\
1.2 & 650.595 & 0.00434022 & 33.8482 \\ \hline
2.1 & 20722.8 & 0.00164722 & 38.9435 \\
2.2 & 23081.0 & 0.00129941 & 44.5132 \\ \hline
3.1 & 3215.05 & 0.00162916 & 13.9127 \\
3.2 & 3356.28 & 0.00188156 & 17.6630 \\ \hline
\end{tabular}
\caption[Maximum, minimum and average absolute errors for $\theta_0$ of different scaling versions over 1000 randomly generated data sets]{Maximum, minimum and average absolute errors for $\theta_0$ of different scaling versions over 1000 randomly generated data sets. Only versions with the same prefix are run on the same data and directly comparable.}
\label{tab:scaling_abs}
\end{table}

The remainder of this section explains the different versions and compares the results:
Version 1.1 is the version discussed in Section \ref{sec:featScal} and implemented in Section \ref{sec:scaling} where $x_{j,conv}$ is computed by $x_{j,conv}=\frac{x_j-avg_j}{max_j}$ with $max_j$ being the maximum of the absolute values of $x_j$.
Version 1.2 on the other hand, computes $x_{j,conv}$ by dividing it by the range of values $x_j$ takes. So the formula is: $x_{j,conv}=\frac{x_j-avg_j}{max_j - min_j}$ with $max_j$ being the maximum of the values without absolute values. Rescaling $\theta_{j,conv}$ to $\theta_j$ uses this range as well: $\theta_{j}=\frac{\theta_{j,conv}}{max_j - min_j}$
As you can see in the tables, version 2 results in bigger errors in both relative errors and absolute errors for $\theta_0$.

Version 2.1 is the same as version 1.1 so $x_{j,conv}=\frac{x_j-avg_j}{max_j}$ holds.
Version 2.2 doesn't use any average values by simply dividing it by the maximum of the absolute value of $x_j$: $x_{j,conv}=\frac{x_j}{max_j}$.
Therefore, $\theta_0$ won't have to be rescaled after gradient descent and all other $\theta$s will be computed in the same way as in version 2.1.
This time the difference in the relative errors is less significant than in comparison 1 and version 2.2 even achieves a slightly lower minimum relative error. However, the average and maximum is still a bit better in version 2.1 and the errors for $\theta_0$ provide the same conclusion.
So in most cases version 2.1 is still preferable. Also worth noting is that on average version 2.2 requires 1000 times more iterations than version 2.1.

Version 3.1 is again the same as version 1.1 and 2.1 where $\theta_{0}$ is computed by subtracting the sum of average values of all of the features from $\theta_{0,conv}$: $\theta_0 = \theta_{0,conv} - \sum_{j=1}^n avg_j$
Version 3.2 on the other hand computes $\theta_0$ by subtracting the sum of the average values of all features from the average value of the label: $\theta_0 = avg_{label} - \sum_{j=1}^n avg_j$
The relative errors for the feature's $\theta$ are obviously the same since nothing changed for them. But in terms of $\theta_0$, version 3.2 performs a bit worse than version 3.1 which also makes this version preferable. However, in Section \ref{sec:psql} a testcase shows that version 3.2 actually performs better for the data used in Section \ref{sec:favorita} which is why most tests in that Section were conducted using version 3.2.
\fi

\begin{figure}[tb]
\centering
\includegraphics[width=\linewidth]{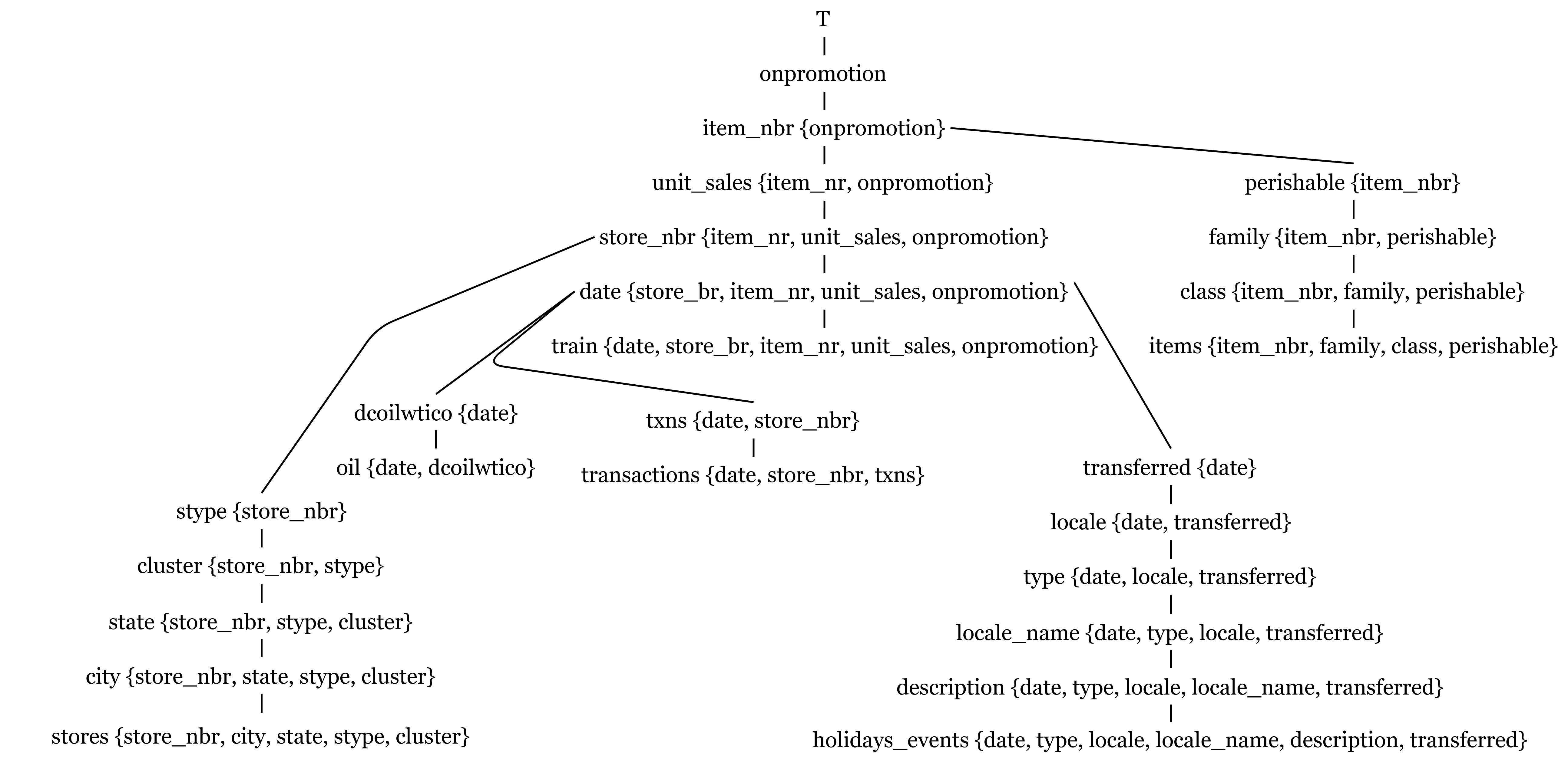}
\caption[Extended variable order used in Section \ref{sec:favorita}]{The extended variable order used in this section (keys of each variable within curly brackets).}
\label{fig:favo_varOr}
\end{figure}

The ``Favorita'' data set \cite{kaggle} is used which represents sales and other information of a grocery chain.
Some of the features had to be converted to numerical values to allow a meaningful linear regression. For example, the column $date$ has been changed to an integer representing $YYYYMMDD - min(YYYYMMDD)$ where the minimum value is subtracted to produce smaller absolute values.
\autoref{fig:favo_varOr} shows the extended variable order used and represents the factorized representation of the natural join of all of relevant relations. 
\noindent
The label, which is predicted in these tests, is $unit\_sales$ and is derived from the features $date$, $store\_nbr$, $item\_nbr$ and $onpromotion$.
All the tests are run on a machine with 220 GiB main memory and two \textit{Intel(R) Xeon(R) CPU E5-2660 v2 @ 2.20GHz} processors resulting in a total number of 20 cores and 40 threads.

All tests are executed on PostgreSQL \cite{psql} v10.10 (\autoref{tab:favo_psql}) and on HyPer~\cite{hyper} (\autoref{tab:favo_hyper}).
The absolute error is calculated by the formula $absErr^{(i)}=\abs{unit\_sales^{(i)} - h_\theta(x^{(i)}}$ and the relative error is determined as $relErr^{(i)} = \frac{absErr^{(i)}}{unit\_sales^{(i)}}$.
\begin{table}[tb]
\caption{Time taken and average errors produced by different implementations. ``Version'' also gives a short description with these abbreviations:\\
$fact$:\quad uses $factorizeSQL(\dots)$ (§\ref{sec:fac}) to precompute cofactors\\
$eps$: \quad \ threshold for gradient descent is set from $10^{-6}$ to $10^{-8}$\\
$\alpha$: \qquad is adjusted differently from the description in §\ref{sec:bgd}\\
$\theta_0$:\qquad $\theta_{0,conv} -\sum avg_j$ is used instead of $avg_{label}-\sum avg_j$\\
$noPre$: aggregates are not precomputed (opposite of $fact$)}
{
\subcaption{PostgreSQL}
\label{tab:favo_psql}
\centering
\begin{tabular}{|l|r|c|c|} \hline
{\textbf{version}} & {\textbf{runtime}} & \textbf{avg. abs. err.} & \textbf{avg. rel. err.} \\ \hline
1 $fact$ & 95m 19s & 8.013579601343 & 2.5518562713374 \\ \hline
2 $noPre$ & 100m 21s & 8.013594277405 & 2.5518638178538 \\ \hline
4 $fact, \alpha$ & 96m 06s & 7.946684605422 & 2.5033955036198 \\ \hline
5 $fact, \alpha,\theta_0$ & 94m 30s & 8.004892391970 & 2.5449049555482 \\ \hline
6 $noPre, \theta_0$ & 101m 46s & 8.054689325103 & 2.5807940338719 \\ \hline
\end{tabular}
\subcaption{HyPer}
\label{tab:favo_hyper}
\begin{tabular}{|l|c|c|c|} \hline
{\textbf{version}} &{\textbf{runtime}} & \textbf{avg. abs. err.} & \textbf{avg. rel. err.} \\ \hline
1 $fact$ & 1m 38s & 7.90330530017246 & 2.4709516975089 \\ \hline
2 $noPre$ & 5m 41s & 7.90332387409202 & 2.4709646443091 \\ \hline
3 $fact, eps$ & 1m 36s & 7.90330179473297 & 2.4709492573143 \\ \hline
4 $fact, \alpha$ & 1m 38s & 7.76919971905285 & 2.3681165862928 \\ \hline
5 $fact, \alpha, \theta_0$ & 1m 39s & 7.94772901833209 & 2.5010221770641 \\ \hline
6 $noPre, \theta_0$ & 5m 37s & 8.04578631749736 & 2.5726866191708 \\ \hline
\end{tabular}
}
\end{table}

\if false
Version 1 uses the $factorizeSQL(\dots)$ (Section \ref{sec:fac}) function to precompute cofactors and uses those for gradient descent as described in Section \ref{sec:bgd}. However, instead of the adjustment to $alpha$ described in that section, it uses a different heuristic which provides more accuracy at the cost of more iterations.



Version 4 doesn't precompute aggregates using $factorizeSQL(\dots)$ and therefore also doesn't use a cofactor matrix for gradient descent.
Instead it simply calculates $\frac{\partial}{\partial\theta_j}\Epsilon(\theta)$ on the materialized join. This means it doesn't take advantage of factorization.
Since this makes it important to reduce the number of iterations, the $alpha$ adjustment described in Section \ref{sec:bgd} is used instead (An execution with the same adjustment mechanism as in version 1-3 was manually aborted after 60 hours since it is totally unfeasible).
Additionally, each iteration of gradient descent is parallelised using OpenMP \cite{openmp} so all aggregates for different $j$s can be computed at the same time.
The results show that this version performs worse than all of the previous versions both in respect to required time and accuracy.


Version 6 uses the factorized representation again, but this time it uses the same adjustment for $alpha$ as version 4 and 5.
The transactions have been split up to separate data manipulation and data definition languages.
The results show that the change of $alpha$ calculation results in an accuracy almost as bad as the non factorized version. However, due to the compatibility changes, this version gained a significant speedup making it faster than the non factorized version 5.

Version 7 didn't result in a notable speedup.
Since the join is materialized in a new table before running gradient descent and all other transactions were already separated from this definition, this isn't too surprising either.
Version 8 takes other adjustment mechanism,m which results in a runtime that is still low while decreasing the error significantly.
Version 9 and 10 both calculate $\theta_0$ by transforming the value of $\theta_{0,conv}$ instead of using the average value of $unit\_sales$.
In both cases this actually leads to the worst accuracy so far without impacting the runtime.

%

This section will reuse some of the tests from Section \ref{sec:psql}.
However, this time the data is located on a HyPer \cite{hyper} server which achieves the results shown in table \ref{tab:favo_hyper}.
\fi
Version 2 and 6 compute gradient descent without factorization, the others with precomputed cofactors.
Version 1 uses the adjustment algorithm for $alpha$ that leads to less iteration but better accuracy.
This version corresponds to the decription in Section \ref{sec:impl}.
Version 2 is equivalent to version 1 but without using $factorizeSQL(\dots)$ to precompute cofactors.
This leads to an decrease in accuracy while at the same time reducing the performance by a considerable amount.
Version 3 equals version 1 but the threshold, which $batchGradientDescent(\dots)$ uses to determine the accuracy, is changed from $10^{-6}$ to $10^{-8}$.
As expected, this leads to a slightly improved result without impacting the time requirement.
Version 4 differs from version 1 since it uses the alternative mechanism for adjusting $alpha$.
Therefore, this leads to an even more accurate result without any performance loss.
Version 5 and 6 correspond to version 4 and 2 but calculate $\theta_0$ by transforming the value of $\theta_{0,conv}$ instead of using the average value of $unit\_sales$. 
This leads to a huge error without having any advantage concerning the runtime.

\pgfplotsset{width=\linewidth,compat=1.9}
\begin{figure}[tb]
\begin{subfigure}[t]{0.49\linewidth}
\centering
\begin{tikzpicture}
\begin{axis}[
	x tick label style={
		/pgf/number format/1000 sep=},
	xtick=data,
	symbolic x coords={1,2,4,5,6,0},
	xlabel=  (HyPer) version number,
	ylabel=Time in Min.,
	enlargelimits=0.05,
	legend style={at={(0.5,-0.26)},anchor=north,legend columns=-1},
	ybar interval=0.7,
]
\addplot 
	coordinates {(1,95.3167) (2,100.35) (4,96.1) (5,94.5) (6,101.7667) (0,0)};
\addplot 
	coordinates {(1,1.6333) (2,5.6833) (4,1.6333) (5,1.65) (6,5.6167) (0,0)};
\legend{PostgreSQL, HyPer}
\end{axis}
\end{tikzpicture}
\end{subfigure}\hfill
\begin{subfigure}[t]{0.49\linewidth}
\centering
\begin{tikzpicture}
\begin{axis}[
	x tick label style={
		/pgf/number format/1000 sep=},
	xtick=data,
	symbolic x coords={1,2,4,5,6,0},
	xlabel=  (HyPer) version number,
	ymin=7.7,
	ylabel= avg. abs. error,
	enlargelimits=0.05,
	legend style={at={(0.5,-0.26)},anchor=north,legend columns=-1},
	ybar interval=0.7,
]
\addplot 
	coordinates {(1, 8.013579601343) (2,8.013594277405) (4,7.946684605422) (5,8.004892391970) (6,8.054689325103) (0,7.7)};
\addplot 
	coordinates {(1,7.90330530017246) (2,7.90332387409202) (4,7.76919971905285) (5,7.94772901833209) (6,8.04578631749736) (0,7.7)};
\legend{PostgreSQL, HyPer}
\end{axis}
\end{tikzpicture}
\end{subfigure}
\caption{Runtime and average error comparison of PostgreSQL and HyPer}
\label{fig:hyper_psql_comp}
\end{figure}
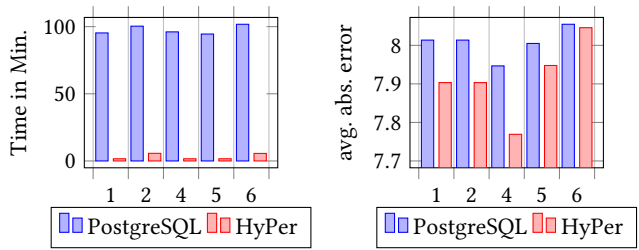

\autoref{fig:hyper_psql_comp} visualizes the differences between PostgreSQL \cite{psql} and HyPer \cite{hyper} in terms of their runtime and average absolute error.
Factorization increased the runtime by 70\% in HyPer and 5\% in PostgreSQL.
HyPer performed almost 50 times faster for the factorized executions and 20 times faster for non-factorized ones than PostgreSQL.
Version 4 produces the most accurate results for both versions but it is also the version where HyPer \cite{hyper} has the biggest advantage over PostgreSQL \cite{psql}.

\section{Conclusion} \label{sec:Concl}
This paper has presented the implementation of factorized learning on top of in-memory database systems.
We first described the concept of factorized learning with the variable order derived from the query plan.
We continued with gradient descent for linear regression on factorized joins.
The paper then presented the implementation, which emits SQL code for feature scaling, cofactor computation, and gradient descent.
We benchmarked our implementation on database systems, with HyPer executing the queries faster than PostgreSQL.
HyPer benefits much more from the factorized representation compared to the materialized version.

For future work, it may be interesting to extend the implementation to polynomial regression.
The added complexity increases the gain from factorized representations even more.
Further tests on different data sets of different sizes might also provide more information about the usefulness of this implementation.


{
\bibliographystyle{ACM-Reference-Format}
\bibliography{literature}
}

\end{document}